\documentclass[aip,reprint]{revtex4-1}

\usepackage{graphicx}
\input{epsf}

\begin{document}


\title{Weakening of the spin density wave gap at low temperatures in SrFe$_{2}$As$_{2}$ single crystals}

\author{Anirban Dutta}
\email[]{adatta@iitk.ac.in}
\affiliation{Department of Physics, Indian Institute of Technology
Kanpur, Kanpur 208016, India}

\author{Neeraj Kumar}
\affiliation{Department of Condensed Matter Physics and Materials
Science, Tata Institute of Fundamental Research, Homi Bhabha Road,
Colaba, Mumbai 400 005, India}

\author{A. Thamizhavel}
\affiliation{Department of Condensed Matter Physics and Materials
Science, Tata Institute of Fundamental Research, Homi Bhabha Road,
Colaba, Mumbai 400 005, India}

\author{Anjan K. Gupta}
\affiliation{Department of Physics, Indian Institute of Technology
Kanpur, Kanpur 208016, India}

\date{\today}

\begin{abstract}
We report on temperature dependent scanning tunneling microscopy and spectroscopy studies of undoped SrFe$_{2}$As$_{2}$ single crystals from 6 K to 292 K. Resistivity data show spin density wave (SDW) transition at T$_{SDW}$ $\approx$ 205 K and the superconducting transition at $\sim$ 21 K while magnetic susceptibility does not show any superconductivity. Conductance maps and local tunneling spectra show an electronically homogeneous surface at all studied temperatures. Tunnel spectra correlate well with the SDW transition with a depression in the density of states near the Fermi energy below T$_{SDW}$. On further lowering the temperature, the width of this partial energy gap associated with the SDW decreases. Using the anti-correlation of superconducting phase with SDW, we interpret this gap weakening as a precursor to superconductivity. This may give rise to a facile nucleation of superconductivity near defects.
\end{abstract}


\maketitle
The superconductivity (SC) in FeAs based layered compounds \cite{kamihara}, occurs in close proximity with the stripe type anti-ferromagnetic (AFM) phase \cite{Johnston-rev,stewart,kordyuk} when the latter is suppressed by chemical doping \cite{Johnston-rev,stewart,kordyuk} or pressure \cite{stewart,chu}. In this respect, pnictides are similar to the layered cuprates and thus the spin-ordering, common to the two, is suspected to have some role in their superconductivity. Thus probing this interface between magnetic ordering and superconductivity has been of significant interest. The parent compounds of pnictides are bad metals, while parent cuprates are antiferromagnetic Mott insulators. The parent compounds of 122 pnictides ($AFe_2As_2$, A: Ca, Ba, Sr or Eu) undergo a spin density wave (SDW) transition at a temperature ($T_{SDW}$) ranging from 125 to 210 K, depending on A \cite{rotter,ronning,krellner,jeevan}. Angle resolved photo-emission spectroscopy (ARPES) shows reconstruction of the energy bands upon entering into the SDW state \cite{yang,jong,zhang,liu}. Some ARPES studies show an energy gap \cite{yang,jong}, while others do not show any gap \cite{zhang,liu} in the SDW state. However, optical conductivity \cite{hu,charnukha,pfuner,wu}, quantum oscillation \cite{sebastian,analytis} and scanning tunneling microscopy and spectroscopy (STM/S) \cite{anirban} studies show an energy gap below the SDW transition. Parts of the Fermi surface disappear due to the nesting of the hole and electron bands at the Fermi energy in the
SDW state and this leads to a reduction in DOS in certain energy range. In pnictides, both SDW and SC originate from the same Fe d-bands. This causes a competition between them over sharing of electronic states common to both gaps resulting in to an anti-correlation between the strengths of the SDW and SC order \cite{pratt,rafael,cai}.

We present STM/S study of
{\it in situ} cleaved SrFe$_2$As$_2$ single crystals between 292 K
and 6 K. Resistivity measurement shows SDW transition at T$_{SDW}$ = 206 K in all crystals and SC transition at T$_c$ = 21 K in some of these undoped crystals. However, no diamagnetic behavior, expected in SC phase, is found. Tunneling spectra show a partial energy-gap below T$_{SDW}$, which becomes more pronounced with cooling but the width of the gap starts reducing near T$_C$. We interpret this weakening of the SDW order at low temperatures as a precursor effect to SC. This could imply an intrinsic propensity of this composition to becoming a SC resulting in to facile nucleation of SC near defects.

Single crystals of SrFe$_2$As$_2$ were grown by the high temperature
solution growth method using Sn-flux under identical conditions as
mentioned in ref. \cite{neeraj}. Electrical resistivity
measurements were done using a standard four-probe method in a
closed cycle refrigerator system. DC magnetization was measured
using a vibrating sample magnetometer (VSM). STM/S studies were done
in a cryogenic vacuum system using a homemade variable temperature
STM system on {\it in situ} cleaved single crystals between 6 and 292 K temperature. Standard ac-modulation
technique was used for STS measurements with a modulation amplitude
of 5 mV and frequency 2731 Hz. The {\it in situ} cleaving was done at room
temperature and at 3 $\times$ 10$^{-6}$ mbar pressure before transferring the crystal to the STM head at low temperature. From tunnel spectra at different locations over 2 $\times$ 2 $\mu$m$^2$ area and conductance maps we found the surface to be electronically homogeneous. Thus we plot spatial average of more than hundred spectra at a particular temperature. Similar temperature dependent spectra on several
pieces of crystals were observed. We plot normalized differential conductance, i.e.
(dI/dV)/($\overline{I/V}$) to sharpen the spectral features where
\begin{eqnarray}
\label{AC} \overline{I/V} &=& \int_{-\infty}^{\infty}
\frac{I(V')}{V'} \exp(\frac{-|V' - V|}{\Delta V})dV'. \nonumber
\end{eqnarray}
We have chosen $\Delta$V
as 10 mV. Feenstra et al. \cite{feenstra}
originally used this technique to study the energy gap of
semiconductors by STM. Temperature also smears the spectral
features of width less than a few k$_B$T. This normalization
procedure sharpens the features that are not thermally smeared by
finite temperature.

\begin{figure}
\includegraphics[width=1.66in]{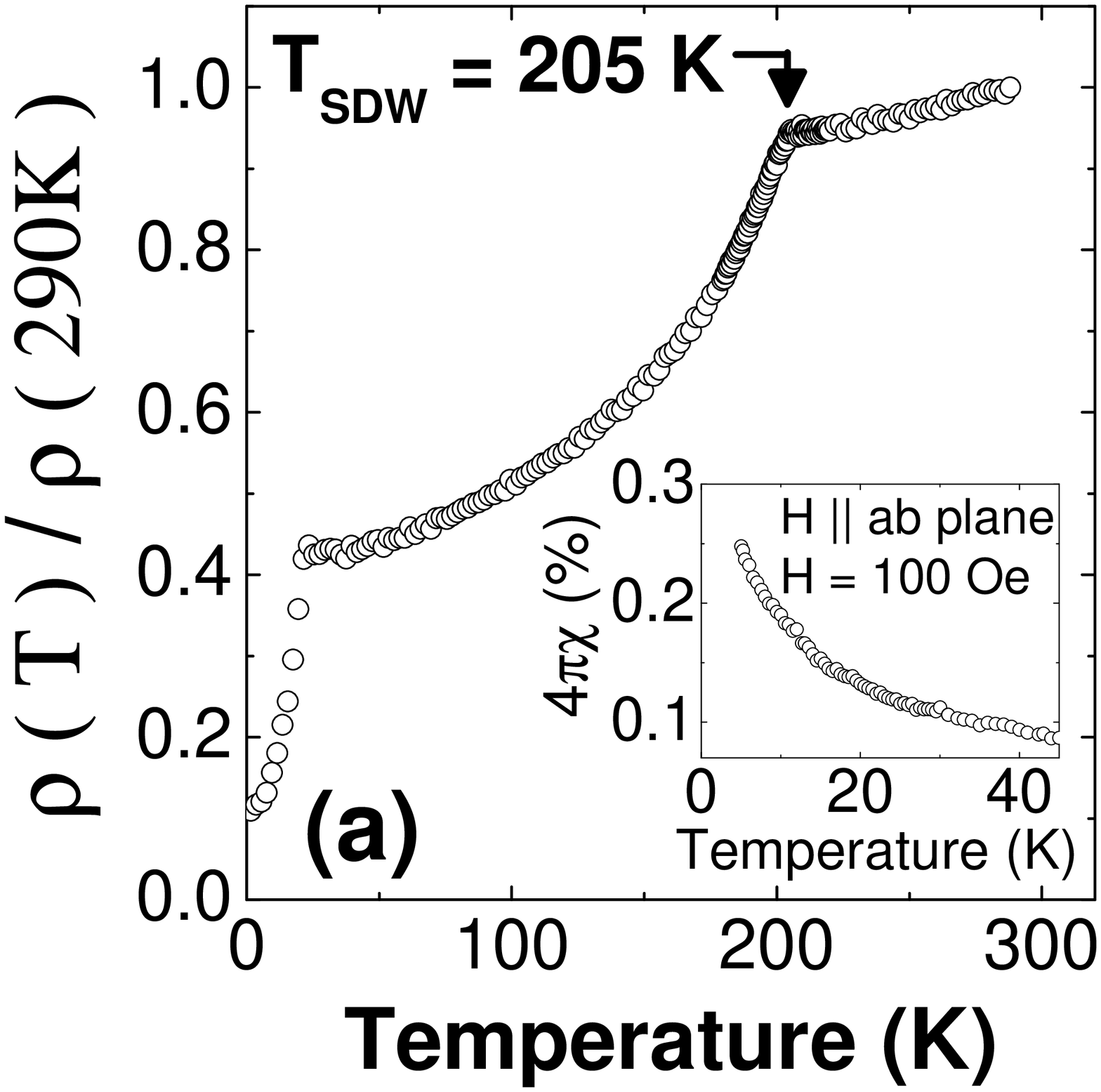}
\includegraphics[width=1.65in]{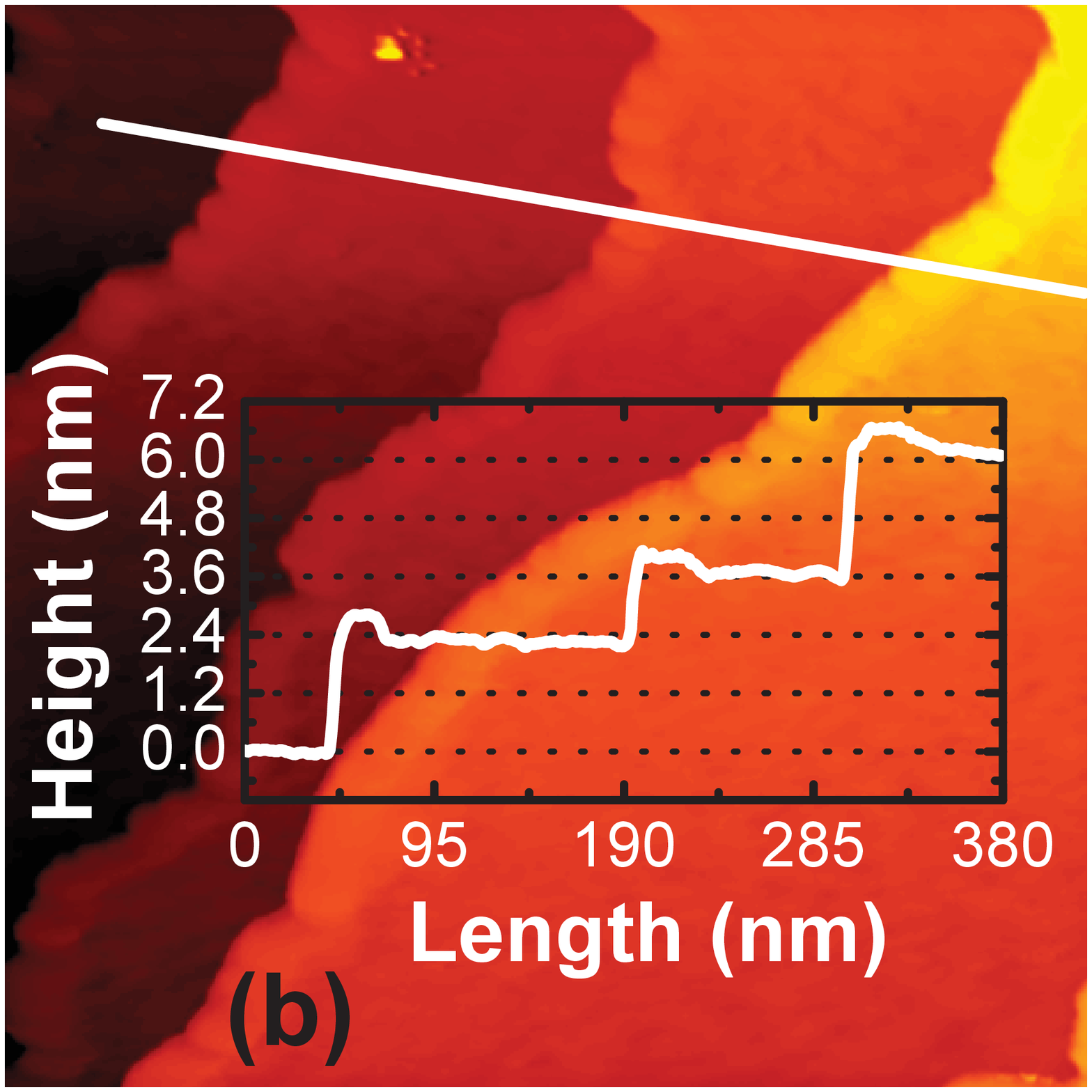}
\caption{\label{fig:resistivity}(Color online) (a) Temperature
dependence of the in-plane electrical resistivity for
SrFe$_{2}$As$_{2}$ single crystal. Inset shows the temperature
dependence of the zero field cooling (ZFC) magnetic susceptibility
for SrFe$_{2}$As$_{2}$ single crystal in 100 Oe. (b) STM topographic
image of ${\emph{in situ}}$ cleaved SrFe$_{2}$As$_{2}$ single
crystal of area 414 $\times$ 441 nm$^2$ at 209.5 K. Image was taken
with 100 pA tunnel current and 400 mV bias. Inset shows the
topographic profile along the marked line.}
\end{figure}

The in-plane resistivity of SrFe$_{2}$As$_{2}$ crystals
is plotted in \mbox{Fig.} \ref{fig:resistivity}(a) as a function of
temperature. The resistivity decreases slowly with
decreasing temperature below room temperature with a sharp kink at around 205 K due to an SDW transition \cite{krellner}. Resistivity decreases rapidly as the
temperature goes down further. The SDW transition opens a gap leading to some loss in density of states (DOS) at the Fermi energy leading to a sharp change in
resistivity behavior. However, rapid decrease of resistivity with
cooling implies only a partial gap and a finite DOS at E$_F$ in the SDW state. Near 21 K, there is sharp drop in resistivity, which we attribute to superconductivity. Such superconducting behavior is usually not seen in undoped pnictides. Saha $\emph{et al.}$ \cite{saha} reported similar behavior together with a diamagnetism below 21 K in some of the as-grown SrFe$_2$As$_2$ crystals confirming the superconductivity. They attributed this superconductivity to the crystallographic strain. The magnetic susceptibility of one of our crystal as plotted in the inset of
\mbox{Fig.} \ref{fig:resistivity}(a) does not show any diamagnetic
behavior below 21 K. Moreover, the resistivity never goes to zero down to 1.5 K.
This implies that a very small portion of the crystal becomes
superconductor below 21 K, which may arise due to SC near defects. Superconducting volume here is small but it is well connected so as to show a drop in resistance. It is possible that this SC portion segregates near the surface or edges.

\begin{figure}
\includegraphics[width=3.4in]{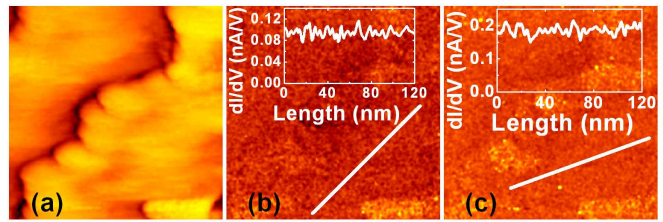}
\caption{\label{fig:condmap-178.5K} (Color online) Simultaneously
acquired STM topographic image and conductance maps of ${\emph{in
situ}}$ cleaved SrFe$_{2}$As$_{2}$ single crystal at 178.5 K of area
190 $\times$ 190 nm$^2$. Images were taken with a tunnel junction of
100 pA tunnel current and 400 mV bias. (a) Topographic image. (b)
Zero-bias conductance map. (c) conductance map at 150 mV. Insets of
(b) and (c) show the dI/dV variations along the marked line in (b)
and (c) respectively.}
\end{figure}

The STM topographic image of SrFe$_{2}$As$_{2}$ single crystal at 209.5 K in \mbox{Fig.} \ref{fig:resistivity}(b) shows step-terrace morphology. The line profile in the inset of \mbox{Fig.} \ref{fig:resistivity}(b) shows that the terraces
are separated by monatomic steps of height 1.2 ($\pm$0.1) nm. The
rms roughness over a terrace is less than 0.1 nm. Similar step-terrace morphology is observed at all studied temperatures, indicating good quality of the cleaved
surface. \mbox{Fig.} \ref{fig:condmap-178.5K} shows the simultaneously
acquired topographic image and conductance maps at 178.5 K. Conductance maps were taken at two different bias voltages: 0 and 150 mV. dI/dV along the marked lines, in the
conductance maps as plotted in the insets of the respective maps, show very little
variation. Similar homogeneous conductance maps were observed at other studied temperature.

\begin{figure}
\includegraphics[width=3.4in]{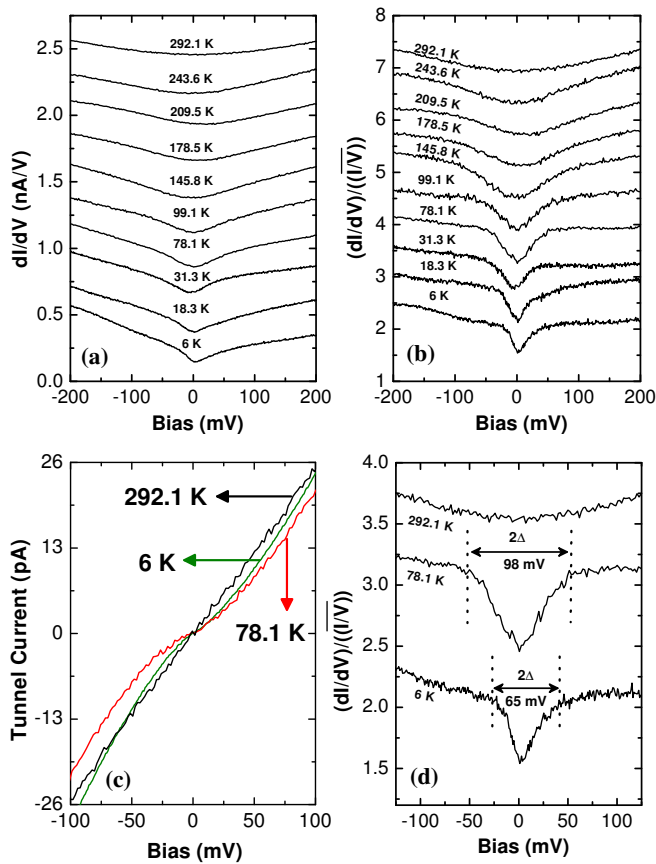}
\caption{\label{fig:spectrancurrent} (Color online) (a) dI/dV Vs. V at 292.1 K - 6 K, (b) normalized
dI/dV Vs. V at 292.1 K - 6 K, (c) Voltage dependent
tunnel current at 292.1 K, 78.1 K and 6.0 K and (d) Expended view of normalized dI/dV Vs. V at 292.1 K, 78.1 K and 6.0 K  of ${\emph{in situ}}$ cleaved SrFe$_{2}$As$_{2}$ single crystals. The tunnel current and spectra were taken
with a junction bias of 300 mV and a tunnel current of 100 pA.
Consecutive spectra in (a), (b) and (d) have been shifted uniformly upwards for clarity.}
\end{figure}

\begin{figure}[h]
\includegraphics[width=3.4in]{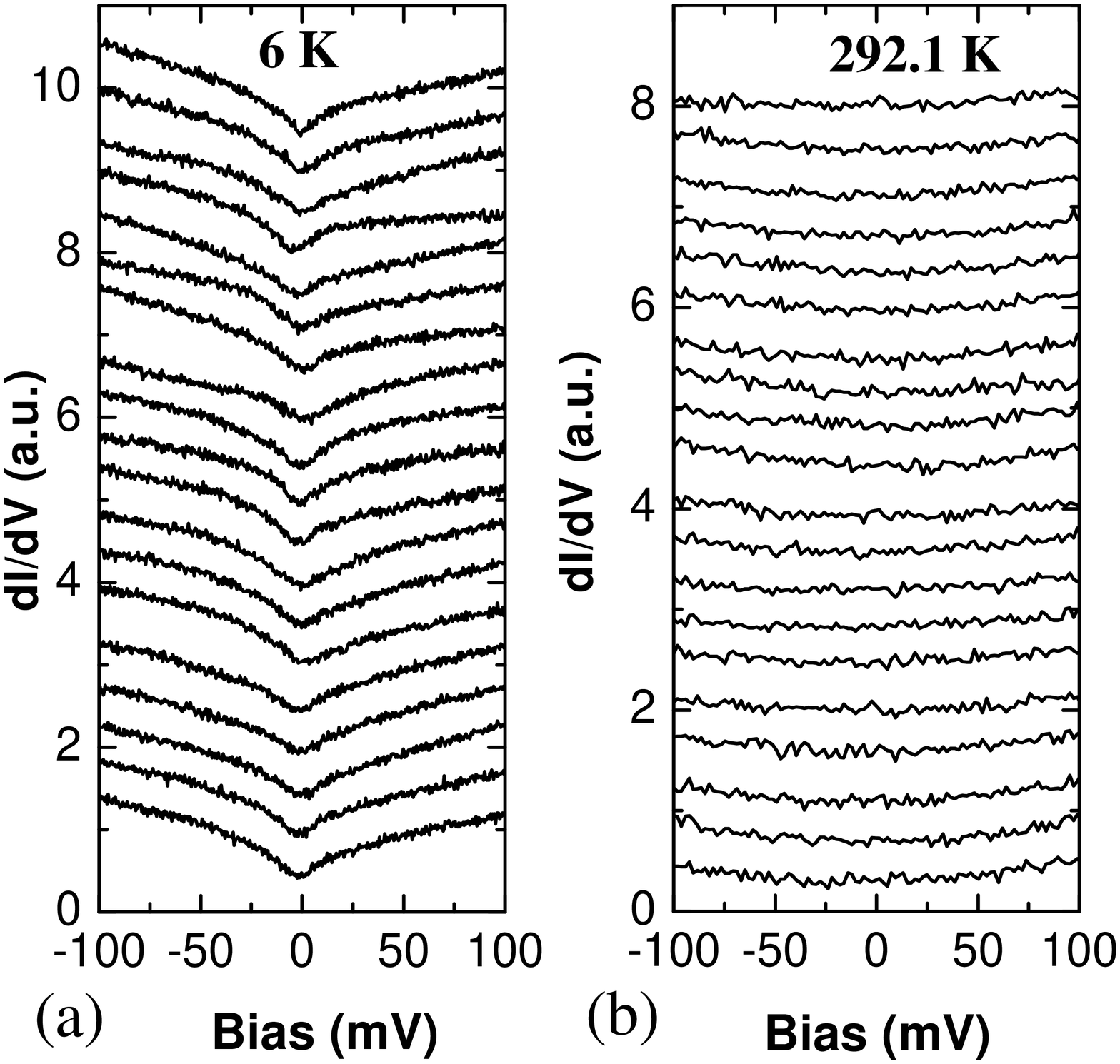}
\caption{\label{fig:spectra-6n292K} dI/dV Vs. V spectra taken at different locations over an area of 2$\times$2 $\mu$m$^2$ of ${\emph{in situ}}$ cleaved SrFe$_{2}$As$_{2}$ single crystal at (a) 6.0 K and (b) 292.1 K. Consecutive spectra have been shifted uniformly upwards for clarity.}
\end{figure}

\mbox{Fig.} \ref{fig:spectrancurrent}(a) shows the temperature dependent spatially averaged tunnel spectra. Three representative I-V spectra at three different studied temperatures are also shown in \mbox{Fig.}
\ref{fig:spectrancurrent}(c). Each plotted conductance spectrum at a particular temperature is a spatial average of about 300 spectra taken over 2 $\times$ 2 $\mu$m$^2$ area. The low temperature tunnel spectra show a dip near zero bias indicating the presence of a gap, while that at
higher temperatures has only a noticeable curvature. Normalized spectra are shown in figure \mbox{Fig.} \ref{fig:spectrancurrent}(b). The normalized spectrum at room temperature does not shows any depression near the E$_F$. But as the temperature goes down, a broad depression arises in the normalized spectra near E$_F$. The broad depression becomes more pronounced as the temperature goes below T$_{SDW}$. We attribute this to the opening of a gap at E$_F$ upon entering the SDW state. The slope of the I-V curve at zero bias which is proportional to the DOS at the E$_F$, is less in the SDW phase at 6 K as compared to the paramagnetic phase at 292.1 K (see the \mbox{Fig.} \ref{fig:spectrancurrent}(c)). The slope at zero-bias is non-zero at 6 K. Thus the gap is not fully open in the SDW phase. Raw spectra taken at different locations over an area of 2$\times$2 $\mu$m$^2$ on the cleaved surface at two extreme temperatures 6 and 292.1 K are shown in the \mbox{Fig.} \ref{fig:spectra-6n292K}. Spectra at different locations have very small variations at both the temperatures and similar kind of local spectra were also observed at all the studied temperatures.

Normalized spectra show a partial gap in the SDW state. With cooling, the gap becomes more pronounced with a gap width of $\sim$ 98 mV at 78 K . The gap width is estimated from the separation between the two edges in the normalized spectra, where the depression starts to appear as marked by the dotted lines in \mbox{Fig.} \ref{fig:spectrancurrent} (d). But with further cooling, the spectra from 31 K - 6 K show that the gap width reduces significantly without change in depth. At 6 K, the estimated gap width is $\sim$ 65 mV. Similar trend is also clearly visible in the dI/dV - V plot. Normalization procedure just sharpens the gap feature without adding (loosing) any extra (existing) feature. At any finite temperature T, the spectral features are smeared out with a few k$_B$T ($\sim$ 3.5k$_B$T) spread. Thus at low temperatures the smearing induced broadening of the SDW gap will be less, which would also make the gap appear smaller but deeper. The reduction in gap from 78 K to 6 K is $\sim$ 33 meV which is $\sim$ 5k$_B$T but the depth of the gap feature is almost unchanged. So the reduction in the gap width cannot be attributed to reduction in thermal smearing.

It is possible that the surface of the studied crystals differs from the bulk due to the room temperature vacuum cleaving or otherwise \cite{jennifer}.  We have observed similar spectra over large area spanning 2$\times$2 $\mu$m$^2$ of cleaved surface and on several different pieces of crystals cleaved similarly. We can not make any comment on the surface reconstruction or termination plane of the studied surface, as we do not have atomically resolved STM images. One STM/S study on room temperature cleaved BaFe$_{1.86}$Co$_{0.14}$As$_2$ showed that termination plane topographic disorder has very little effect on the low-lying electronic states of these systems \cite{massee}. A hard X-ray photoemission also supports this finding \cite{jong2}. Thus we believe that the observed spectral features near Fermi energy do reflect intrinsic property of these crystals.

Optical spectroscopy on SrFe$_2$As$_2$ \cite{hu,charnukha}, CaFe$_2$As$_2$ \cite{charnukha} and BaFe$_2$As$_2$ \cite{hu,charnukha} shows presence of two gaps. In SrFe$_2$As$_2$, Hu $\emph{et al.}$ \cite{hu} identified two gaps with peaks at 500 cm$^{-1}$ and 1500 cm$^{-1}$ which correspond to 62.04 meV and 168.75 meV respectively. Charnukha $\emph{et al.}$ \cite{charnukha} also reported two gaps for SrFe$_2$As$_2$, but with slightly lower values (380 and 1215 cm$^{-1}$). Our tunneling spectra show only one gap and it's magnitude almost matches with the smaller gap observed by Hu $\emph{et al.}$ \cite{hu}. A pseudogap of 500 cm$^{-1}$ without any sign of gap at higher energy is observed in one optical study \cite{pfuner} on BaFe$_2$As$_2$. Another optical investigation \cite{wu} on EuFe$_2$As$_2$ also doesn't show the second gap. Our earlier tunneling study \cite{anirban} on EuFe$_2$As$_2$ also showed a single gap like spectra.

In 122-pnictides the SDW and SC phases coexist over a finite doping range \cite{Johnston-rev,stewart,kordyuk} and they can coexist both homogeneously and inhomogeneously as reported by several groups \cite{pratt,rafael,cai,julien,park}. The STM/S study by Cai $\emph{et al.}$ \cite{cai} in underdoped NaFe$_{1-x}$Co$_x$As$_2$ showed an anti-correlation between the strength of the SDW and SC order, indicating a competition between the two. Nevertheless, the SC order clearly anti-correlates, particularly in the underdoped regime, with the SDW in the sense that SC appears when SDW declines. Reduction of SDW gap at low temperature in our results could be a precursor effect to an upcoming SC order, which does not quite happen in the bulk of this undoped crystal. Although resistivity shows a SC transition near 21 K in some of our crystals similar to an earlier report \cite{saha} where SC was attributed to strain related to defects. Magnetic fluctuations are believed to play an important role in SC of pnictides but long range magnetic order does not favor SC. The weakening of the SDW order at low temperatures will certainly make it easier for SC to nucleate, at least near defects.

In conclusion, our variable temperature STM/S study shows a partial energy gap due to SDW order below 200 K, which starts becoming weaker at low temperatures below about 30K. The resistivity of some of the crystals shows a partial SC transition at 21 K with negligible SC fraction inferred from magnetic susceptibility. The weakening of SDW order at low temperatures could be a precursor to SC order and indicates an intrinsic inclination of these crystals to becoming a superconductor resulting into facile nucleation of SC near defects as seen in resistivity.

We thank Sourabh Barua for his help in the resistivity measurement. Anirban acknowledges financial support from the CSIR of the Government of India. A.K.G. acknowledges a research grant from the CSIR of the Government of India. We would like to acknowledge Avinash Singh and Sayandip Ghosh for helpful discussions.

\end{document}